\documentclass[aip,
pdftex,
amsmath,
amssymb,
preprint,%
]{revtex4-1}

\usepackage[pdftex]{graphicx}
\usepackage{dcolumn}
\usepackage{bm}
\usepackage[mathlines]{lineno}

\usepackage[utf8]{inputenc}
\usepackage[T1]{fontenc}
\usepackage{mathptmx}
\usepackage{etoolbox}

\def\revVI#1{#1}

\def\vort{\varpi}
\def\pd#1#2{\frac{\partial #1}{\partial #2}}
\def\ptb#1{{#1}_{1}}
\def\equ#1{{#1}_{0}}
\def\ion{\mathrm i}
\def\para{\parallel}
\def\gvis#1#2{\mathcal{G}\left(#1,#2\right)}
\def\curv#1{\mathcal{K}\left(#1\right)}
\def\vave#1{\langle #1 \rangle_{V}}
\def\bracket#1#2{\left[#1, #2\right]}
\def\bvec#1{\bm{#1}}
\makeatletter
\def\@email#1#2{%
 \endgroup
 \patchcmd{\titleblock@produce}
  {\frontmatter@RRAPformat}
  {\frontmatter@RRAPformat{\produce@RRAP{*#1\href{mailto:#2}{#2}}}\frontmatter@RRAPformat}
  {}{}
}%
\makeatother
\begin{document}

\preprint{AIP/123-QED}

\title[Two-stage crash process in RDBM driven ELM crash]{Two-stage Crash 
Process in Resistive Drift Ballooning Mode Driven ELM Crash}

\author{H. Seto}
\email{seto.haruki@qst.go.jp}
\affiliation{National Institutes for Quantum Science and Technology, 
Rokkasho, Aomori 039-3212, Japan}

\author{X.Q. Xu}%
\affiliation{Lawrence Livermore National Laboratory, Livermore, CA 94550, USA}

\author{B.D. Dudson}
\affiliation{Lawrence Livermore National Laboratory, Livermore, CA 94550, USA}

\author{M. Yagi}
\affiliation{National Institutes for Quantum Science and Technology, 
Rokkasho, Aomori 039-3212, Japan}

\date{\today}

\begin{abstract}
 We report a two-stage crash process in edge localized mode (ELM) 
 driven by resistive drift-ballooning modes (RDBMs) numerically 
 simulated in a full annular torus domain. 
 In the early nonlinear phase, the first crash is triggered  by linearly 
 unstable RDBMs and $m/n=2/1$ magnetic islands  are nonlinearly excited via 
 nonlinear couplings of RDBMs. 
 Simultaneously, \revVI{middle-$n$} RDBM turbulence develops but is 
 poloidally localized around X-points of the magnetic islands, leading to 
 the small energy loss. 
 Here $m$ is the poloidal mode number, $n$ is the toroidal mode number, 
 the $q=2$ rational surface exists at the pressure gradient peak, 
 and $q$ is the safety factor, respectively.
 The second crash occurs in the late nonlinear phase. 
 \revVI{Low-$n$} magnetic islands are also excited around the $q=2$ surface 
 via nonlinear couplings among the middle-$n$ turbulence. 
 Since the turbulence develops from the X-points of higher harmonics of 
 $m/n=2/1$ magnetic islands, it expands out poloidally.
 The second crash is triggered when the turbulence covers the whole poloidal 
 region. 
 A scan of toroidal wedge number $N$, where full torus is divided into $N$ 
 segments in the toroidal direction, also reveals that the first crash 
 process becomes more prominent with the higher toroidal wedge number where 
 the RDBMs play a dominant role. 
 These results indicate that nonlinear interactions of all channels in the 
 full torus domain can significantly affect the trigger dynamics of ELMs 
 driven by the RDBMs.
\end{abstract}
\maketitle

\section{Introduction} \label{sec:1} 
The intermittent heat loads by edge localized modes 
(ELMs)~\cite{Zohm:PPCF1996} in H-mode tokamak plasma~\cite{Wagner:PRL1982} 
should be avoided or mitigated below heat load constraints on plasma facing 
components, which is one of key issues for 
ITER~\cite{Kukushkin:FED2011,Gunn:NF2017} and DEMO~\cite{Asakura:NME2021}. 
Nonlinear numerical simulations are powerful tools to understand ELM dynamics 
and to calculate ELM energy loss so that several nonlinear MHD codes 
such as JOREK~\cite{Huysmans:NF2007,Czarny:JCP2008,Hoelzl:NF2021}, 
NIMROD~\cite{Sovinec:JCP2004},  
M3D-C1~\cite{Ferraro:POP2010,Wingen:PPCF2015}, 
MEGA~\cite{Todo:POP1998,Todo:POP2017,RiveroRodriguez:NF2023} and 
BOUT++~\cite{Dudson:CPC2009,Xu:PRL2010,Zhu:CPC2021} have been developed 
and have provided qualitative understanding of ELMs.

BOUT++ code is a plasma fluid simulation framework solving plasma fluid 
equations as initial value problems in arbitrary curvilinear coordinate 
systems with finite difference methods. 
For three dimensional tokamak boundary plasma simulations including ELMs, 
BOUT++ code employs the quasi-ballooning coordinate 
system~\cite{Dudson:CPC2009} consisting of orthogonal flux surface 
coordinates~\cite{Dhaeseleer:BOOK1991} for differences in the radial 
direction and field aligned coordinates~\cite{Beer:POP1995} for differences 
along the equilibrium magnetic field line. 
With this coordinate system, BOUT++ code can calculate middle-$n$ 
$(\mathcal{O}(n) > 1)$ and high-$n$ $(\mathcal{O}(n) \gg 1)$ plasma 
instabilities with reasonable computational cost and high accuracy, 
which is suitable for simulations of ELMs by the ballooning modes.

Its computation domain was however limited to an $1/N$-th annular toroidal 
wedge to remove low-$n$ $(\mathcal{O}(n) \sim 1)$ mode components from the 
system to avoid numerical instabilities.
Here, in the $1/N$-th annular toroidal wedge torus, a full torus domain 
is divided equally into $N$ parts in the toroidal direction. 
This is because the flute-ordering approximation neglecting differences 
along the magnetic field is required in the field solver of flow potential 
in the \revVI{quasi-ballooning} coordinate system.

Recently we have resolved this issue by implementing a hybrid field 
solver~\cite{Seto:CPC2023} consisting of a 2D field solver for $n=0$ 
and low-$n$ modes, and a flute-ordered 1D field solver for moderate-$n$ 
and high-$n$ modes to address ELM crash simulations in the full annular 
torus domain.
Taking the full annular torus domain is important not only 
for simulating ELMs by kink/peeling modes, ELMs with 
resonant magnetic perturbations (RMPs)~\cite{Evans:PRL2004,Orain:POP2013}, 
QH-mode accompanied with low-$n$ edge harmonic 
oscillations~\cite{Burrell:PPCF2002,Liu:NF2015} 
but also for simulating ELMs by ballooning modes, 
ELMs with turbulence transport and so on. 
For example, the number of nonlinear mode-mode couplings in the simulated 
system can change the nonlinear criterion of the ELM crash~\cite{Xi:PRL2014}. 
A full-f core gyrokinetic simulation  reveals that using too large toroidal 
wedge number $N$ can result in the false convergence of turbulence heat 
transport level~\cite{Kim:POP2017}, which may also occur in edge turbulence 
transport simulations.

Our recent work~\cite{Seto:CPC2023} also shows that taking full annular 
torus domain can qualitatively change crash process of the ELMs by the 
\revVI{resistive drift ballooning modes (RDBMs)} compared to that in a quarter 
torus domain with $N=4$, however the crash mechanism has not been clarified 
in detail. 
In this paper, the \revVI{RDBM-driven} ELM in the full annular torus domain 
is analyzed to understand its crash mechanism as well as quantitative 
difference between ELMs with different toroidal wedge numbers.

The rest of this paper is organized as followings. 
\revVI{A} set of governing equation and \revVI{a} MHD equilibrium used for 
ELM crash simulations are described in Sec.\ref{sec:2}.
The crash process in the ELM crash in the full annular torus domain is 
analyzed. 
The crash mechanism is discussed in detail. 
In addition, the impact of the wedge torus domain on ELM crash is also 
discussed in Sec.\ref{sec:3}.
The paper is finally summarized in  Sec.~\ref{sec:4}.

\section{Simulation setup} \label{sec:2} 
The following scale-separated four-field reduced model with the flat ion 
density profile $n_{\rm i} =\bar{n_{\rm i}}$ describing the 
\revVI{RDBM}~\cite{Seto:CPP2020} is employed for ELM crash simulations,
\begin{eqnarray}
 \pd{\ptb{\vort}}{t} &=& 
  - \bracket{\ptb{F}}{\vort} 
  - \bracket{\equ{F}}{\ptb{\vort}} 
  + \gvis{\ptb{P}}{F} 
  + \gvis{\equ{P}}{\ptb{F}} 
  \nonumber \\ &&
  + \curv{\ptb{P}}
  - \equ{B}\partial_{\para}\left(\frac{\ptb{J_{\para}}}{\equ{B}}\right)
  + \equ{B}\bracket{\ptb{A_{\para}}}{\frac{J_{\para}}{\equ{B}}}
  + \mu_{\para}\partial_{\para}^{2}\ptb{\vort}
  + \mu_{\perp}\nabla^{2}_{\perp}\ptb{\vort}\revVI{,} \label{eq:vort}\\
 \pd{\ptb{A_{\parallel}}}{t} &=& 
  - \bracket{\phi}{\ptb{A_{\parallel}}} 
  - \partial_{\para}\ptb{\phi} 
  + \delta\left(\partial_{\para}\ptb{P} - \bracket{\ptb{A_{\para}}}{P} \right) 
  + \eta\ptb{J} - \lambda\nabla_{\perp}^{2}\ptb{J}\revVI{,} \label{eq:ohm}\\
 \pd{\ptb{P}}{t} &=& -\bracket{\phi_{1}}{P} -\bracket{\phi_{0}}{P_{1}} 
  - 2\beta_{*}\curv{\ptb{\phi}}
  \nonumber \\ && 
  - \beta_{*}\equ{B}\partial_{\para}
  \left(\frac{\ptb{v}+d_{\ion}\ptb{J_{\para}}}{\equ{B}}\right)
  + \beta_{*}\equ{B}
  \bracket{\ptb{A_{\para}}}{\frac{\ptb{v}+d_{\ion}\ptb{J_{\para}}}{\equ{B}}}
  +\chi_{\para}\partial_{\para}^{2}\ptb{P}
  +\chi_{\perp}\nabla_{\perp}^{2}\ptb{P}\revVI{,} \label{eq:pres}\\
 \pd{\ptb{v_{\para}}}{t} &=& -\bracket{\phi}{\ptb{v_{\para}}}
 - \frac{1}{2}\partial_{\para}\ptb{P}
 + \frac{1}{2}\bracket{\ptb{A_{\para}}}{P} 
 + \nu_{\perp}\nabla_{\perp}^{2}\ptb{v_{\para}}, \label{eq:vpara} \\
 \vort &=& \nabla_{\perp *}^{2} F,\quad 
 \equ{\phi} = -\delta\equ{P},\quad
 \quad \ptb{J_{\para}} = \nabla _{\perp}^{2} \ptb{A_{\para}},
 \quad \ptb{\bvec{B}} =\nabla \ptb{A_{\para}} \times \equ{\bvec{b}}
 \nonumber \\
 \partial_{\parallel} f &=& \equ{\bvec{b}}\cdot \nabla f,\quad
  \partial_{\parallel}^{2} f =  
  \partial_{\parallel}\left(\partial_{\parallel} f \right),\quad
  \nabla_{\perp} f = \nabla f -\equ{\bvec{b}}\equ{\bvec{b}}\cdot \nabla f,
  \quad \nabla_{\perp}^{2}f = \nabla \cdot \nabla_{\perp} f\revVI{,} 
  \nonumber \\
 \bracket{f}{g} &=& \frac{\equ{\bvec{b}}\times \nabla_{\perp} f \cdot 
  \nabla_{\perp} g}{\equ{B}},\quad  
  \nabla_{\perp *}^{2} f = \nabla \cdot 
  \left(\frac{\nabla_{\perp} f}{\equ{B}^{2}}\right),\quad 
  \curv{f} = \frac{\equ{\bvec{b}}\times \equ{\bvec{\kappa}} 
  \cdot \nabla f}{\equ{B}}\revVI{,}
  \nonumber \\
 \gvis{f}{g} &=& \frac{\delta}{2}
  \left(\bracket{f}{\nabla_{\perp *}^{2}g}
   +\bracket{f}{\nabla_{\perp *}^{2}g}
   +\nabla_{\perp *}^{2}\bracket{f}{g}\right). \nonumber 
\end{eqnarray}
Here $\vort$ is the vorticity defined with the generalized flow potential 
$F=\phi +\delta P$, $P=\equ{P} + \ptb{P}$ is the plasma pressure, 
$\phi = \equ{\phi} + \ptb{\phi}$ is the electrostatic potential, 
$\delta = d_{\rm i}/4$ is the factor for electron and ion diamagnetism 
for the isotropic pressure case $P_{\rm e}=P_{\rm i} =P/2$, 
$d_{\rm i}$ is the ion skin depth, 
$J_{\para} = \equ{J_{\para}}+\ptb{J_{\para}}$ is the parallel current density, 
$A_{\para}=\ptb{A_{\para}}$ is the parallel magnetic potential, 
$\bvec{B} = \equ{\bvec{B}} + \ptb{\bvec{B}}$ is the magnetic field intensity, 
$\mu_{\para}$ is the parallel viscosity for vorticity
$\mu_{\perp}$ is the perpendicular viscosity for vorticity, 
$\eta$ is the resistivity, $\lambda$ is the hyper-resistivity,
$\beta_{*}=\equ{B}^{2}/[0.5+\equ{B}^{2}/(5\equ{P}/3)]$ 
is the compression factor, 
$\equ{\bvec{b}}=\equ{\bvec{B}}/{\equ{B}}$ is the unit vector 
along the equilibrium magnetic field, 
$\equ{\bvec{\kappa}}=\equ{\bvec{b}}\cdot \nabla \equ{\bvec{b}}$ 
is the magnetic curvature, $v_{\para}$ is the ion parallel flow, 
$\chi_{\para}$ is the parallel heat diffusivity,
$\chi_{\perp}$ is the perpendicular heat diffusivity, 
and $\nu_{\perp}$ is the perpendicular viscosity for parallel flow, 
respectively. 
In this model, the subscript ``0'' represents an equilibrium part and the 
subscript ``1'' represents a perturbed part of physical quantities, 
$f(\bvec{x},t) =\equ{f}(\bvec{x}) + \ptb{f}(\bvec{x},t)$.
The ion gyroviscous cancellation is modeled in the Chang-Callen 
manner~\cite{Chang:PoF1992} in the vorticity equation and the set 
of equations is normalized with poloidal Alfv\'{e}n unit with 
the reference length $\bar{R}=3.5~[\mathrm{m}]$, 
the reference magnetic intensity $\bar{B}=2.0~[\mathrm{T}]$, 
the reference ion number density 
$\bar{n_{\rm i}} = 1.0\times 10^{19}~[\mathrm{m}^{-3}]$, 
the deuterium mass and the effective charge number $Z=1.0$. 

In this paper, we employ constant \revVI{resistivities} 
$\eta=1.0\times 10^{-8}$ and $\lambda=1.0 \times 10^{-12}$, and viscosities 
and diffusivities$\mu_{\perp}= \chi_{\perp}= \nu_{\perp}=1.0\times 10^{-7}$ 
and $\mu_{\para}= \chi_{\para}= 1.0 \times 10^{-1}$ as our previous 
works~\cite{Seto:CPC2023, Seto:CPP2020}.
It should be noted that the equilibrium $E\times B$ flow, or the equilibrium 
radial electric field, is modeled to cancel with the equilibrium ion 
diamagnetic flow $\equ{F} =0$ so that the neoclassical poloidal flow and its 
return flow~\cite{Itoh:PPCF1996} are not taken into account.
Modeling the return flow is a key to simulate the bifurcation 
between zonal flow and streamer formation, which is left for future works. 
It should be also noted that the anomalous electron heat 
diffusivity~\cite{Rechester:PRL1978} and the hyper-resistivity by magnetic 
stochastisation~\cite{Kaw:PRL1979} are not taken into account due to the usage 
of the constant hyper-resistivity and the linearized second parallel 
derivative $\partial_{\para}^{2}$ lacking magnetic flatter effects. 
This means that ELM energy loss in this work is driven 
by $E\times B$ convection and numerical diffusion.
These are left for future works.

\begin{figure}[!t]
 \includegraphics[width=0.495\linewidth]{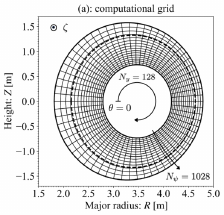}
 \includegraphics[width=0.495\linewidth]{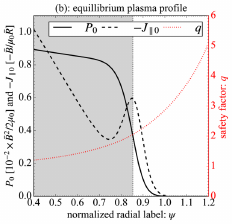}
 \caption{(a): computational grid based on \revVI{the} shifted circular 
 equilibrium and (b): \revVI{its} plasma profiles; 
 the equilibrium plasma pressure $\equ{P}$ (black solid), 
 the equilibrium parallel current density $\equ{J_{\para}}$ on the outer 
 mid-plane (black dashed) and \revVI{the} safety factor $q$ (red dotted), 
 respectively.}
 \label{fig:01}
\end{figure}

The computational grid for the orthogonal flux surface coordinate system 
$(\psi,\theta,\zeta)$ and plasma profiles used in the ELM crash 
simulations are shown in Fig.~\ref{fig:01}.
Here the $\psi$ is the poloidal flux function, $\theta$ is the orthogonal 
poloidal angle and $\zeta$ is the geometrical toroidal angle, respectively. 
The computational grid is constructed from a shifted circular equilibrium 
marginally unstable against ideal ballooning 
mode~\cite{Wilson:PoP2002,Snyder:PoP2002} generated by TOQ equilibrium
code~\cite{Miller:NF1987,Burke:PoP2010}. 
The quasi-ballooning coordinate system~\cite{Dudson:CPC2009} consists of 
the flux surface coordinate system $(\psi,\theta,\zeta)$ and 
the field-aligned coordinate system $(x,y,z)$, 
where $x=\psi$ is the radial label, $y=\theta$ is the parallel label, and
$z=\zeta-\alpha$ is the binormal label with the shift angle 
$\alpha = \int_{\pi}^{\theta}(\equ{\bvec{B}}\cdot\nabla\zeta/
\equ{\bvec{B}}\cdot \nabla\theta)d\theta$, respectively. 
The coordinate transform between the flux surface coordinate system
and the field-aligned coordinate system is performed in the Fourier space 
with respect to the toroidal mode number $n$ using the phase relation 
$z=\zeta-\alpha$, which is briefly reviewed in Ref.~\cite{Seto:PoP2019}. 

\begin{figure}[!t]
 \includegraphics[width=0.5\linewidth]{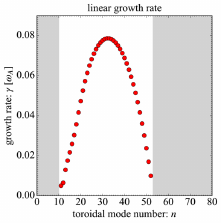}
 \caption{linear growth rate of RDBM instability, where the shaded regions 
 $0 \le n \le 10$ and $n \ge 53$ are stable against RDBM instability.} 
 \label{fig:02}
\end{figure}

The linear growth rate of RDBMs for this equilibrium is shown in 
Fig.~\ref{fig:02}.
The largest growth rate is given by $\gamma/\omega_{A} = 7.86\times 10^{-2}$ 
for the toroidal mode number $n=32$.
It is found that \revVI{the RDBMs are stable for $n \le 10$ and $n \ge 53$}.

For ELM crash simulations, the number of radial grids 
$N_{\psi}$ is 1028 for $0.4 \le \psi \le 1.2$ and the 
number of parallel grids (or poloidal grids) $N_{y}$ is 128.
The number of binormal grids (or toroidal grids) $N_{z}$ is $256/N$ 
for $0\le z <2\pi/N$, where the grid width in the binormal direction 
is kept constant for different toroidal wedge numbers.
Here the radial and poloidal grid \revVI{resolutions} are fine enough, 
which is discussed in section 5.2 in Ref.~\cite{Seto:CPC2023}.
In the hybrid field solver calculating the generalized flow 
potential~\cite{Seto:CPC2023}, 
$0 \le n \le 4$ mode components are calculated by the 2D field solver and 
$4 < n \le 80$ mode components are calculated by the flute-ordered 1D field 
solver, where $n \ge 81$ components are removed with a low-pass filter.
In all ELM crash simulations reported in this work, initial perturbations are 
set on all modes except $n=0$ mode to introduce nonlinear 
couplings~\cite{Xi:PRL2014} self-consistently. 
The set of radial boundary conditions is 
the Neumann boundary condition $\partial_{\psi} f =0$ 
at the inner boundary $\psi=0.4$ 
and Dirichlet boundary condition $f =0$ at the outer boundary  $\psi=1.2$ for 
$f= \ptb{\vort}, \ptb{P}, \ptb{A_{\para}}, \ptb{v_{\para}}, \ptb{F}, 
\ptb{J_{\para}}$.

\section{Two-stage crash process in RDBM-driven ELM crash} \label{sec:3}
In the first part of this section, the crash process in the RDBM-driven 
ELM crash in the full annular torus domain with $N=1$~\cite{Seto:CPC2023} 
is analyzed in detail.
In the second part of this section, the dependence of toroidal wedge numbers 
with $N=1,2,4$ on the ELM crash process is investigated to understand the 
qualitative difference of them.

\begin{figure}[!t]
 \includegraphics[width=0.5\linewidth]{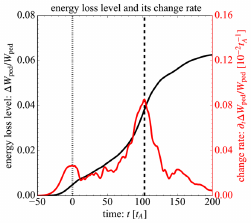}
 \caption{time evolution of ELM energy loss level 
 $\Delta W_{\rm ped}/W_{\rm ped}$ with the black solid curve and the left 
 vertical axis and that of change rate of ELM energy loss level 
 $\partial_{t} \Delta W_{\rm ped}/W_{\rm ped}$ with the red solid curve and 
 the right vertical axis in the full torus case.
 Here the first crash at $t=0t_{A}$ is highlighted with the black dotted line 
 and the second crash at $t=103t_{A}$ is highlighted with the black dashed 
 line, respectively.}
 \label{fig:03}
\end{figure}

The time evolution of the ELM energy loss level and its change rate in the 
full torus case are summarized in Fig.~\ref{fig:03}, where the time label is 
set to be $t=0t_{A}$ at the time when the $n=32$ component of the 
perpendicular kinetic energy gets saturated. 
Here, the energy loss level $\Delta W_{\rm ped}/W_{\rm ped}$ is defined 
by the ratio of the energy lost from the region inside the rational surface 
of the initial pressure gradient peak $V_{\rm ped}$ highlighted with 
the black dotted line and shaded area in Fig.~\ref{fig:01}(b), 
\begin{eqnarray}
 \Delta W_{\rm ped}/W_{\rm ped} = -\int_{V_{\rm ped}} \ptb{P}dV 
  / \int_{V_{\rm ped}} \equ{P} dV.
\end{eqnarray}
It is clear that the change rate of the energy loss level has the two peaks 
at $t=0t_{A}$ and $t=103t_{A}$. 
Hereafter we define the first peak at $t=0t_{A}$ to be the first crash and 
the second peak at $t=103t_{A}$ as the second crash, respectively. 

Equations of the system energies in this model are defined as the followings.
The equation of $n=n'$ component of volume-averaged perpendicular kinetic 
energy $W_{\rm k}(t,n')$ can be derived from vorticity equation 
\revVI{Eq.~(\ref{eq:vort})} by multiplying with $n=n'$ component of the 
generalized flow potential $\ptb{F}^{n=n'}$ and taking its volume average 
over the computation domain $V$, 
\begin{eqnarray}
 \pd{}{t}W_{\rm k}(t,n') &=& 
  T_{\rm k, RS}^{\rm nl}(t,n') 
  + T_{\rm k, CV}^{\rm li}(t,n') + T_{\rm k, LB}^{\rm li}(t,n') 
  + T_{\rm k, MS}^{\rm nl}(t,n') + T_{\rm k, VD}^{\rm li}(t,n'), 
  \label{eq:Wk}
\end{eqnarray}
with
\begin{eqnarray}
 W_{\rm k}(t,n') &=& \vave{\frac{1}{2\equ{B}^{2}} 
 \left|\nabla_{\perp}\ptb{F}^{n=n'}\right|^{2}},
 \label{eq:WkEn}
\end{eqnarray}
and
\begin{eqnarray}
 T_{\rm k, RS}^{\rm nl}(t,n') &=& \vave{\ptb{F}^{n=n'}\bracket{\ptb{F}}{\vort}}
 + \vave{\ptb{F}^{n=n'}\bracket{\equ{F}}{\ptb{\vort}}} 
 - \vave{\ptb{F}^{n=n'}\gvis{\ptb{P}}{F}} 
 - \vave{\ptb{F}^{n=n'}\gvis{\equ{P}}{\ptb{F}} }, \\
 T_{\rm k, CV}^{\rm li}(t,n') &=& - \vave{\ptb{F}^{n=n'}\curv{\ptb{P}}}, \\
 T_{\rm k, LB}^{\rm li}(t,n') &=& \vave{\ptb{F}^{n=n'}
  \equ{B}\partial_{\para}\left(\frac{\ptb{J_{\para}}}{\equ{B}}\right)},
\end{eqnarray}
\begin{eqnarray}
 T_{\rm k, MS}^{\rm nl}(t,n') &=& - \vave{\ptb{F}^{n=n'}
 \equ{B}\bracket{\ptb{A_{\para}}}{\frac{J_{\para}}{\equ{B}}}},\label{eq:Tms}\\
 T_{\rm k, VD}^{\rm li}(t,n') &=& 
 - \vave{\ptb{F}^{n=n'}\mu_{\para}\partial_{\para}^{2}\ptb{\vort}}
 - \vave{\ptb{F}^{n=n'}\mu_{\perp}\nabla^{2}_{\perp}\ptb{\vort}}, 
\end{eqnarray}
where 
$T_{\rm k, RS}^{\rm nl}(t,n')$ is the contribution from Reynolds stress 
terms, 
$T_{\rm k, CV}^{\rm li}(t,n')$ is the contribution from geodesic curvature 
term, 
$T_{\rm k, LB}^{\rm li}(t,n')$ is the contribution from line-bending term,
$T_{\rm k, MS}^{\rm nl}(t,n')$ is the contribution from Maxwell stress term, 
$T_{\rm k, VD}^{\rm li}(t,n')$ is the energy loss by numerical viscosity 
terms, and $\vave{f} =V^{-1}\int_{V}f dV$ is the volume average operation, 
respectively. 
The equation of $n=n'$ component of volume-averaged magnetic energy 
$W_{\rm m}(t,n')$ can be also derived from Ohm's law 
\revVI{Eq.~(\ref{eq:ohm})} as, 
\begin{eqnarray}
 \pd{}{t}W_{\rm m}(t,n')  
  &=&  T_{\rm m, EH}^{\rm li}(t,n')  
  +  T_{\rm m, EH}^{\rm nl}(t,n')
  + T_{\rm m, RD}^{\rm li}(t,n'), \label{eq:Wm}
\end{eqnarray}
with
\begin{eqnarray}
 W_{\rm m}(t,n') &=&  \vave{\frac{1}{2} 
 \left|\nabla_{\perp} \ptb{A_{\para}}^{n=n'}\right|^{2}},
\label{eq:WmEn}
\end{eqnarray}
and
\begin{eqnarray}
 T_{\rm m, EH}^{\rm li}(t,n') &=& \vave{\ptb{J_{\para}}^{n=n'} 
  \partial_{\para}\left(\ptb{\phi}-\delta\ptb{P}\right)}, \\
 T_{\rm m, EH}^{\rm nl}(t,n') &=&  -\vave{\ptb{J_{\para}}^{n=n'} 
  \bracket{\ptb{A_{\para}}}{\phi-\delta P}}, \\
 T_{\rm m, RD}^{\rm li}(t,n') &=& -\vave{\ptb{J_{\para}}^{n=n'} 
  \eta\ptb{J_{\para}}}
  - \vave{ \lambda\left|\nabla_{\perp}\ptb{J_{\para}}^{n=n'}\right|^{2}},
\end{eqnarray}
where $T_{\rm m, EH}^{\rm li}(t,n')$ and $T_{\rm m, EH}^{\rm nl}(t,n')$
are the linear and nonlinear contribution from electrostatic potential 
and electron Hall terms, 
$T_{\rm m, RD}^{\rm li}(t,n')$ is the energy loss 
by the resistive dissipation, respectively. 
The equation of $n=n'$ component of internal energy $W_{\rm p}(t,n')$ can be 
also derived from equation of pressure \revVI{Eq.~(\ref{eq:pres})}, 
\begin{eqnarray}
 \pd{}{t}W_{\rm p}(t,n') &=&  T_{\rm p, PV}^{\rm nl}(t,n') 
  + T_{\rm p, CO}^{}(t,n') + T_{\rm p, PD}^{\rm li}(t,n')\revVI{,} 
  \label{eq:Wp}
\end{eqnarray}
with
\begin{eqnarray}
 W_{\rm p}(t,n') &=& \vave{\frac{1}{4\beta_{*}}
\left|\ptb{P}^{n=n'}\right|^{2}}\revVI{,} \label{eq:WpEn}
\end{eqnarray}
and
\begin{eqnarray}
 T_{\rm p, PV}^{\rm nl}(t,n') &=& 
 - \vave{\frac{1}{2\beta_{*}}\ptb{P}^{n=n'}\bracket{\phi_{1}}{P}} 
 - \vave{\frac{1}{2\beta_{*}}\ptb{P}^{n=n'}
 \bracket{\phi_{0}}{P_{1}}},\revVI{,} \label{eq:Tpv} \\
 T_{\rm p, CO}^{}(t,n') &=& T_{\rm p, CE}^{\rm li}(t,n')
 + T_{\rm p, CJ}^{\rm li}(t,n') + T_{\rm p, CJ}^{\rm nl}(t,n')
 + T_{\rm p, CV}^{\rm li}(t,n') + T_{\rm p, CV}^{\rm nl}(t,n')\revVI{,} \\
 T_{\rm p, CE}^{\rm li}(t,n') &=& -\vave{\ptb{P}^{n=n'}\curv{\ptb{\phi}}},\\
 T_{\rm p, CJ}^{\rm li}(t,n') &=& 
 -2\delta\vave{\ptb{P}^{n=n'}\equ{B}\partial_{\para}
 \left(\frac{\ptb{J_{\para}}}{\equ{B}}\right)},
\end{eqnarray}
\begin{eqnarray}
 T_{\rm p, CJ}^{\rm nl}(t,n') &=& 
 2\delta\vave{\ptb{P}^{n=n'} \equ{B}
 \bracket{\ptb{A_{\para}}}{\frac{\ptb{J_{\para}}}{\equ{B}}}}, \\
 T_{\rm p, CV}^{\rm li}(t,n') &=& 
 - \frac{1}{2}\vave{\ptb{P}^{n=n'}\equ{B}\partial_{\para}
 \left(\frac{\ptb{v}}{\equ{B}}\right)}\revVI{,}\\ 
 T_{\rm p, CV}^{\rm nl}(t,n') &=& \frac{1}{2}\vave{\ptb{P}^{n=n'}\equ{B} 
  \bracket{\ptb{A_{\para}}}{\frac{\ptb{v}}{\equ{B}}}},\\  
 T_{\rm p, HD}^{\rm li}(t,n') &=&
 \vave{\frac{\ptb{P}^{n=n'}}{2\beta_{*}}\chi_{\para} 
 \partial_{\para}^{2}\ptb{P}} 
 + \vave{\frac{\ptb{P}^{n=n'}}{2\beta_{*}}\chi_{\perp}
 \nabla_{\perp}^{2}\ptb{P}},
\end{eqnarray}
where $T_{\rm p, PV}^{\rm nl}(t,n')$ is the contribution from \revVI{the} 
$E\times B$ convection terms, 
$T_{\rm p, CO}^{}(t,n')$ is the contribution from the compression terms,
$T_{\rm p, CE}^{\rm li}(t,n')$ is the contribution from the $E\times B$ flow 
compression term,
$T_{\rm p, CJ}^{\rm li}(t,n')$ and $T_{\rm p, CJ}^{\rm nl}(t,n')$ are 
the contributions from linear and nonlinear part of parallel current 
compression terms, 
$T_{\rm p, CV}^{\rm li}(t,n')$ and $T_{\rm p, CV}^{\rm nl}(t,n')$ are 
the contributions from linear and nonlinear part of parallel ion flow 
compression terms, and $T_{\rm p, HD}^{\rm li}(t,n')$ is energy loss by 
numerical diffusion terms, respectively.
Finally, the equation of $n=n'$ mode component of parallel kinetic energy 
$W_{\rm v}(t,n') = \vave{\frac{1}{2}{\ptb{v_{\para}}^{n=n'}}^{2}}$ can be 
also derived from equation of ion parallel flow by multiplying with $n=n'$ 
component of ion parallel flow $\ptb{v_{\para}}^{n=n'}$ 
\revVI{Eq.~(\ref{eq:vpara})} and averaging it over the computation domain.
Its expression is \revVI{however} not shown here since the parallel kinetic 
energy is only weakly coupled with the internal energy and its budget is not 
analyzed in this work. 

\begin{figure}[!hbt]
\includegraphics[width=0.5\linewidth]{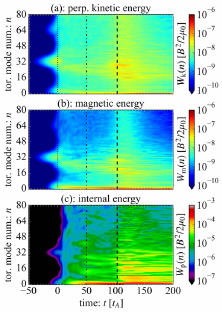}
 \caption{time evolution of energy spectra; 
 (a): the perpendicular kinetic energy,
 (b): the magnetic energy and (c): the internal energy in the full torus case, 
 where the dotted lines are at $t=0t_{A}$, 
 the dash-dot lines are at an intermediate time between the two crashes 
 $t=50t_{A}$, and the dashed lines are at $t=103t_{A}$, respectively.}
 \label{fig:04} 
 \includegraphics[width=0.95\linewidth]{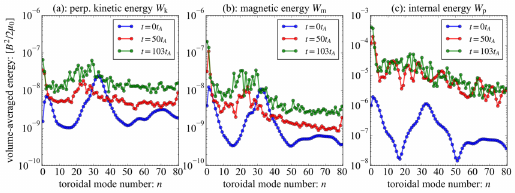}
 \caption{
 toroidal mode spectra of the three system energies at $t=0t_{A}$ (blue), 
 $t=50t_{A}$ (red), and $t=103t_{A}$ (green); 
 (a) the perpendicular kinetic energy, (b) the magnetic energy, 
 (c) the internal energy, respectively.}
 \label{fig:05}
\end{figure}

The time evolution of toroidal mode spectra of the three system energies 
Eqs.~(\ref{eq:WkEn}), (\ref{eq:WmEn}) and (\ref{eq:WpEn}) \revVI{is} 
summarized in Fig.~\ref{fig:04}. 
It is found that the RDBMs whose peak is given by $n=32$ mode directly trigger 
the first crash. 
At the first crash, energy cascades to higher toroidal modes and 
inverse energy cascades to lower toroidal modes than $n=32$ occur. 
The latter contributes to the spectrum peak shifts from $n=32$ to lower 
toroidal modes.
On the other hand, the $n=0$ and low-$n$ components of system energies 
grow around the first crash. 
At the second crash, $n=0$ and $n=1$ component are comparable to down-shifted 
middle-$n$ peak in the perpendicular kinetic energy and are dominant 
components in the other system energies, which is clearly seen in the time 
slices of toroidal mode spectra of the \revVI{three} system energies in 
Fig~\ref{fig:05}.

\begin{figure}
 \includegraphics[width=0.5\linewidth]{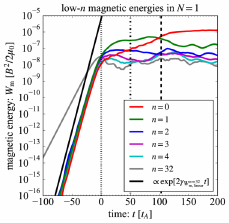} 
 \caption{time evolution of low toroidal mode number components of magnetic 
 energy in the full torus case, where the black dotted line is at $t=0t_{A}$, 
 the black dash-dot line is at $t=50t_{A}$, the black dashed line is at 
 $t=103t_{A}$, the gray curve is the time evolution of the initially most 
 unstable $n=32$ component of magnetic energy, and the black sold line 
 represents a curve with twice larger growth rate than that of the $n=32$ one.}
 \label{fig:06} 
\end{figure}

For generation mechanism of low-$n$ modes, time evolution of low-$n$ magnetic 
energies \revVI{is} shown in Fig.~\ref{fig:06}.
In the early nonlinear phase before the first crash, \revVI{the} $n=0$ and 
low-$n$ magnetic energies develop with the growth rate almost twice of that 
of \revVI{the} $n=32$ magnetic energy.
This indicates that \revVI{the} $n=0$ and low-$n$ magnetic energies are 
driven by nonlinear couplings among the RDBMs.
After the first crash, the low-$n$ magnetic energies get saturated while the 
$n=0$ magnetic energy nonlinearly grows until the second crash\revVI{, and 
the $n=0$ magnetic energy} becomes a dominant component after the second crash.
A detailed discussion on the generation mechanism of low-$n$ 
\revVI{components of the magnetic energy} after the first crash is given later 
with three-wave analyses of nonlinear terms.

\begin{figure}
 \includegraphics[width=0.95\linewidth]{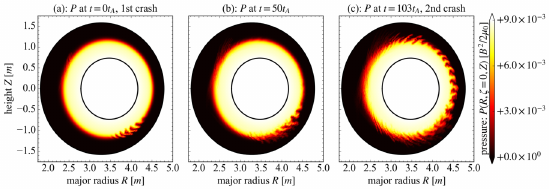}
 \caption{pressure profiles on $\zeta=0$-plane in cylindrical coordinates 
 at (a): $t=0t_{A}$, (b): $t=50t_{A}$, and (c): $t=103t_{A}$, respectively.}
 \label{fig:07}
\end{figure}

Figure~\ref{fig:07} shows the time evolution of pressure profiles on the 
$\zeta=0$ plane.
At the first crash, fine scale pressure fluctuations driven by the RDBMs are 
poloidally localized in the two regions, upper left and lower right regions on 
the $q=2$ flux surface where the initial pressure gradient peak exists.  
After the first crash, pressure fluctuations expand out poloidally and finally 
cover over the flux surface \revVI{at the second crash}. 
This spatial structure can be related with $m/n=2/1$ magnetic fluctuations.
Energy budgets of $n=1$ system energies are analyzed to clarify generation 
mechanism of $n=1$ magnetic fluctuations and magnetic field line tracing 
analyses are carried out to clarify the role of magnetic field fluctuations 
on ELM energy loss process.

\begin{figure}
 \includegraphics[width=0.5\linewidth]{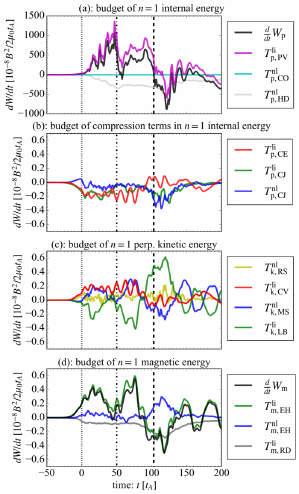}
 \caption{Time evolution of 
 (a) the budget of $n=1$ internal energy equation,
 (b) the budget of the compression terms coupling with 
 the perpendicular kinetic energy and the magnetic energy
 in $n=1$ internal energy equation, (c) the budget of $n=1$ perpendicular 
 kinetic energy, and (d) the budget of $n=1$ magnetic energy, respectively. 
 Here, \revVI{the energy loss by numerical viscosity terms 
 $T_{\rm k, VD}^{\rm li}$ and the change rate of $n=1$ kinetic energy 
 $\frac{d}{dt}W_{\rm k}$ in Fig.~\ref{fig:08}(c)} have not been plotted 
 for readability, and the black dotted lines are $t=0t_{A}$, the black 
 dash-dot lines are $t=50t_{A}$, and the black dashed lines are $t=103t_{A}$, 
 respectively.}
 \label{fig:08} 
\end{figure}

The budget of $n=1$ internal energy Eq.~(\ref{eq:Wp}) is summarized in 
Fig.~\ref{fig:08}(a). 
Here the terms with \revVI{the} superscript ``li'' transfer energy within 
$n=1$ components and the terms with \revVI{the} superscript ``nl'' transfer 
energy among all toroidal mode components, respectively.
The $n=1$ internal energy is driven by \revVI{the} $E\times B$ convection 
term $T_{\rm p, PV}^{\rm nl}$ and its change rate becomes deeply negative 
after the second crash.
The sum of contributions from \revVI{the} compression terms $T_{\rm p, CO}$ 
is a higher order term and is much smaller than the others.
The compression terms however form energy channels between the other system 
energies and should be kept for self-consistent energy transfer in the 
simulated system.

The energy transfers between the $n=1$ internal energy $W_{\rm p}(t,n=1)$, 
perpendicular kinetic energy $W_{\rm k}(t,n=1)$ and magnetic energy 
$W_{\rm m}(t,n=1)$ are summarized in Fig.~\ref{fig:08}(b)-(d),  
where the terms with the same color construct energy transfer channels.
The budget of $n=1$ parallel kinetic energy $W_{\rm v}(t,n=1)$ has not be 
shown since the parallel kinetic energy is weakly coupled only with the 
internal energy via parallel pressure compression terms 
$T_{\rm p, CV}^{\rm li}$ and $T_{\rm p, CV}^{\rm nl}$, 
and has little impact on the magnetic energy generation in this model. 

For the $n=1$ magnetic energy generation, the contribution from \revVI{the} 
$E\times B$ convection in \revVI{the $n=1$} internal energy equation 
$T_{\rm p, PV}^{\rm nl}$ and that from \revVI{the} Maxwell stress in the 
$n=1$ perpendicular kinetic energy equation $T_{\rm k, MS}^{\rm nl}$ are 
positive after the first crash while the other nonlinear contributions are 
small till the second crash. 
Here the positive and negative \revVI{signs indicate} the energy gain and 
loss, respectively.
It should be noted that nonlinear couplings by means of Poisson brackets in 
reduced MHD model generate the tearing parity components via nonlinear parity 
mixing~\cite{Sato:POP2017,Ishizawa:PPCF2019}.
The $n=1$ magnetic energy is mainly driven by the linear part of 
\revVI{the} electrostatic and electron Hall term $T_{\rm m, EH}^{\rm li}$ 
which cannot contribute to the parity mixing directly, 
but the $n=1$ pressure and electrostatic potential perturbations 
are driven by $T_{\rm p,PV}^{\rm nl}$ and $T_{\rm k,MS}^{\rm nl}$ which 
generate the tearing-parity components.
Then, the tearing parity of \revVI{the $n=1$ parallel magnetic potential} 
is given by the linear combination of pressure and electrostatic potential 
\revVI{via} $T_{\rm m, EH}^{\rm li}$.
This is consistent with the exponential growth of \revVI{the} $n=1$ magnetic 
energy with \revVI{the} growth rate almost twice larger than that of 
\revVI{the} $n=32$ magnetic energy at the early nonlinear phase.
Our previous work~\cite{Seto:CPC2023} also confirmed that 
the $n=1$ tearing parity is obtained after the second crash.

\begin{figure}[b]
 \includegraphics[width=0.95\linewidth]{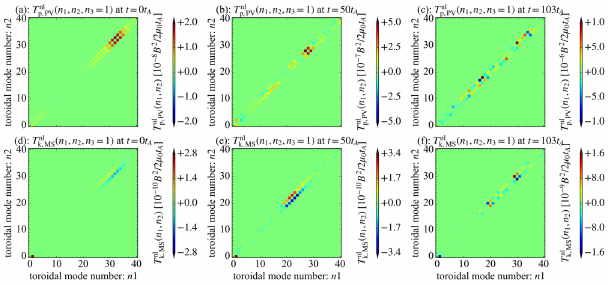}
 \caption{ 
 Three-wave analyses of the $E\times B$ convection term in the $n=1$ internal 
 energy equation at (a): $t=0t_{A}$, (b): $t=50t_{A}$ and (c): $t=103t_{A}$, 
 and three-wave analyses of the Maxwell stress term in the $n=1$ perpendicular 
 kinetic energy equation at (d): $t=0t_{A}$, (e): $t=50t_{A}$ and (f): 
 $t=103t_{A}$, respectively.}
 \label{fig:09} 
\end{figure}

To clarify the nonlinear energy transfer channel driving the $n=1$ 
magnetic energy, three-wave analysis on the $E\times B$ convection term 
$T_{\rm p, PV}^{\rm nl}(t,n_{1},n_{2},n_{3})$ and that on the Maxwell 
stress term $T_{\rm k, MS}^{\rm nl}(t,n_{1},n_{2},n_{3})$,
\begin{eqnarray}
 T_{\rm p, PV}^{\rm nl}(t,n_{1},n_{2},n_{3}) &=& 
  - \vave{\frac{1}{2\beta_{*}}\ptb{P}^{n=n_{3}}
  \bracket{\phi^{n=n_{1}}}{P^{n=n_{2}}}}\revVI{,} \\
 T_{\rm k, MS}^{\rm nl}(t,n_{1},n_{2},n_{3}) &=& 
  - \vave{\ptb{F}^{n=n_{3}}\equ{B} 
  \bracket{\ptb{A_{\para}}^{n=n_{1}}}{\frac{J_{\para}^{n=n_{2}}}{\equ{B}}}},
\end{eqnarray}
are applied with $n_{3}=1$ at $t=0t_{A}$, $t=50t_{A}$ and $t=103t_{A}$.
Here, pairs of $n_{1}$ and $n_{2}$ satisfying $|n_{1}-n_{2}|=1$ can 
contribute \revVI{to} $n=1$ energy \revVI{gains} and the results are 
summarized in Fig.~\ref{fig:09}.

For the $E\times B$ convection term, pairs in the RDBMs have positive 
contributions to the $n=1$ internal energy at the first crash and pairs in the 
down-shifted middle-$n$ turbulence also have positive contributions to the 
$n=1$ internal energy at $t=50t_{A}$. 
At the second crash, a wide range of spectrum have positive and 
negative \revVI{contributions} to the $n=1$ internal energy.
For the Maxwell stress, at the first crash and $t=50t_{A}$, 
the coupling with $(n_{1},n_{2})=(1,0)$ is the dominant positive contribution 
so that the $n=1$ perpendicular kinetic energy gets energy from $n=0$ and 
$n=1$ magnetic energy. 
Pairs in middle-$n$ modes satisfying $n_{1}-n_{2} =+1$ have negative 
\revVI{contributions} and those satisfying $n_{1}-n_{2} = -1$ have positive 
contributions at the first crash and $t=50t_{A}$.
At the second crash, the coupling with $(n_{1},n_{2})=(1,0)$ has \revVI{the} 
large negative contribution so that the $n=1$ perpendicular kinetic energy 
gives energy to \revVI{the} $n=0$ and $n=1$ magnetic \revVI{energies}.

\begin{figure}
 \includegraphics[width=0.95\linewidth]{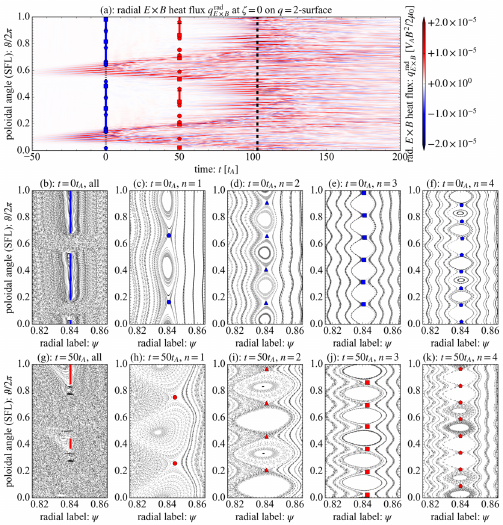}
 \caption{
 (a): spatio-temporal structure of radial heat flux by $E\times B$ convection 
 on the $q=2$ flux surface, Poincare plots of magnetic field line at 
 $t=0t_{A}$ including 
 (b): all toroidal mode numbers of magnetic perturbations, 
 (c): $n=1$ single mode,  (d): $n=2$ single mode, 
 (e): $n=3$ single mode,  (f): $n=4$ single mode, 
 and those at $t=50t_{A}$ including 
 (g): all toroidal mode numbers of magnetic perturbations, 
 (h): $n=1$ single mode, (i): $n=2$ single mode, 
 (j): $n=3$ single mode, (k): $n=4$ single mode, respectively.
 Here the solid lines indicate the existing area of magnetic islands, and
 the circles, the triangles, the squares and the pentagons are the X-points 
 of $m/n=2/1$, $m/n=4/2$, $m/n=6/3$, and $m/n=8/4$ magnetic islands 
 at $t=0t_{A}$ (blue) and $t=50t_{A}$ (red), respectively.}
\label{fig:11}
\end{figure}

To understand the generation mechanism of the poloidally localized pressure 
fluctuations at the first crash in Fig.~\ref{fig:07}, spatio-temporal 
structure of the radial $E\times B$ heat flux $q^{\rm rad}_{E\times B}$ along 
$\zeta=0$ line on the $q=2$ flux surface and time evolution of Poincare plot 
of magnetic field lines are calculated, 
which are summarized in Fig.~\ref{fig:11}.
Here the straight-field-line \revVI{(SFL)} poloidal angle $\vartheta$ in 
Fig.~\ref{fig:11} is defined with the magnetic local pitch and the orthogonal 
poloidal angle $\theta$
\begin{eqnarray}
 \vartheta \equiv =\frac{1}{q}\int_{0}^{\theta}
  \frac{\equ{\bvec{B}} \cdot \nabla\zeta}{\equ{\bvec{B}} \cdot \nabla\theta} 
  d\theta,
\end{eqnarray}
and the radial $E\times B$ heat flux $q^{\rm rad}_{E\times B}$ is also 
defined with the radial $E\times B$ flow $v^{\rm rad}_{E\times B}$,
\begin{eqnarray}
 v_{E\times B}^{\rm rad} &=& 
  \bvec{v}_{E\times B}
  \cdot h_{\psi}\nabla \psi 
  = \frac{1}{B_{p}R}\pd{\phi}{z} -\frac{B_{t}}{B^{2}h_{\theta}}
  \pd{\phi}{y}\revVI{,} \\
 q_{E \times B}^{\rm rad} &=& P{v}_{E\times B}^{\rm rad} 
  = P\left( \frac{1}{B_{p}R}\pd{\phi}{z} 
      -\frac{B_{t}}{B^{2}h_{\theta}}\pd{\phi}{y}\right).
\end{eqnarray}

Figure~\ref{fig:11} clearly shows that the $n=1$ magnetic perturbation forms 
$m/n=2/1$ magnetic islands at the first crash and the radial heat flux only 
exists the region with \revVI{the} stochastic magnetic field around 
the X-points of \revVI{the} $m/n=2/1$ magnetic islands.
Here, the width of \revVI{the} stochastic region at the first crash seems 
to be determined by the location of X-points of $m/n=2/1$ and $m/n=8/4$ 
magnetic islands.
The magnetic topology gets more stochastic with time since the higher 
\revVI{harmonics of the $m/n=2/1$} magnetic islands develop on the $q=2$ 
rational surface and \revVI{the} magnetic island overlapping strongly enhance magnetic 
stochastisation\cite{Diamond:POF1984}.
It is should be noted that the radial heat flux flows out from the X-point of 
higher harmonics of $m/n=2/1$ magnetic islands, which enhances energy loss 
level. 
In this simulation, the radial heat flux flows out from two X-points of 
$m/n=8/4$ magnetic island around $\vartheta/2\pi \sim 0.34~\mbox{and}~0.84$ 
at $t=50t_{A}$.

\begin{figure}
 \includegraphics[width=0.95\linewidth]{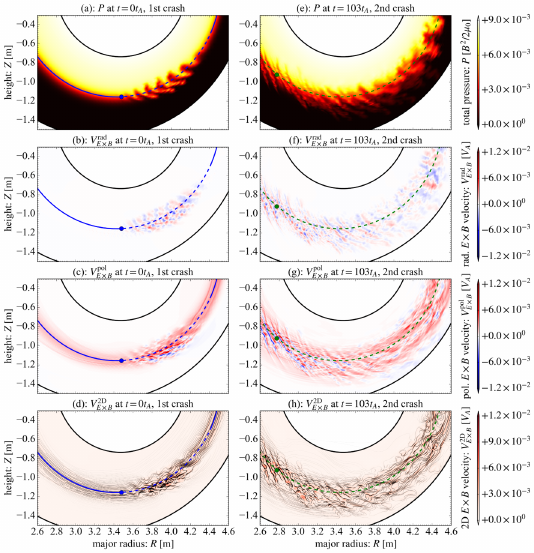}
 \caption{poloidal slices of (a) total pressure, (b) radial $E\times B$ flow, 
 (c) poloidal $E\times B$ flow and (d) 2D velocity of $E\times B$ flow 
 on the $\zeta=0$ plane at the first crash, and 
 those of (e) total pressure, (f) radial $E\times B$ flow, 
 (g) poloidal $E\times B$ flow and (h) 2D velocity of $E\times B$ flow 
 on the $\zeta=0$ plane at the second crash, respectively. 
 Here the dashed blue and green curves are the $q=2$ surface, the blue and 
 green circles are the X-points of $m/n=2/1$ magnetic islands in the case 
 where only $n=1$ magnetic perturbation is taken into account, and the solid
 blue curves are the region where the magnetic islands exist, respectively.}  
\label{fig:12} 
\end{figure}

The poloidal plot of total pressure and $E \times B$ flow components around 
the low \revVI{magnetic} field side of \revVI{the} X-points of \revVI{the} 
$m/n=2/1$ magnetic islands at the first and second crashes are summarized 
in Fig.~\ref{fig:12}.
Here the magnetic islands are shown with \revVI{the} solid 
curves\revVI{, and the poloidal $E\times B$ flow $v_{E\times B}^{\rm pol}$} 
and the 2D $E\times B$ flow $v_{E\times B}^{\rm 2D}$ are given by
\begin{eqnarray}
 v_{E \times B}^{\rm pol} &=& 
  \bvec{v}_{E\times B}\cdot h_{\theta} \nabla \theta 
  = \frac{B_{t}B_{p}R}{\equ{B}^{2}} \pd{\phi}{\psi}\revVI{,} \\
 v_{E \times B}^{\rm 2D} &=& 
  \sqrt{(v_{E\times B}^{\rm rad})^{2} + (v_{E\times B}^{\rm pol})^{2}}.
\end{eqnarray}

At the first crash, the radial flow filaments are localized in the region 
without magnetic islands and the poloidal flow consisting of the small scale 
zonal flow and the large scale mean flow forms a laminar structure in the 
region with magnetic islands. 
The vortex flow pattern enhancing energy loss is therefore observed in the 
region without magnetic islands. 
At the second crash when the magnetic topology around the $q=2$ flux surface 
is fully stochastic, the pressure filaments expand poloidally as well as the 
radial flow filaments. 
The vortex pattern of 2D $E\times B$ flow therefore is formed over the domain 
and results in the increase of ELM energy loss.
These results are consistent with the pressure profiles and the 
spatio-temporal analysis of radial $E\times B$ heat flux in Fig.~\ref{fig:11}. 
Here, an analysis of the causality between the magnetic stochastisation and 
the vortex flow pattern formation is left for a future work.

\begin{figure}
\includegraphics[width=0.5\linewidth]{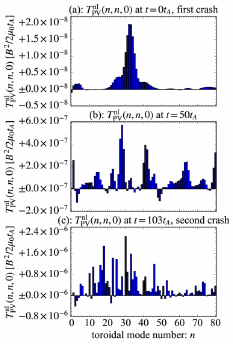}
 \caption{ three-wave analysis of the $E\times B$ convection term in $n=0$ 
 internal energy at (a): $t=0t_{A}$, (b): $t=50t_{A}$, and 
 (c): $t=103t_{A}$, respectively.}
 \label{fig:14}
\end{figure}

To clarify the trigger mechanism of two crashes, the three-wave analysis of 
\revVI{the} $E\times B$ convection term in the $n=0$ internal energy 
$T^{\rm nl}_{\rm p,PV}(t,n_{1}=n,n_{2}=n,n_{3}=0)$ is investigated, which is 
summarized in Fig.~\ref{fig:14}. 
Here the positive contribution \revVI{enhances} the $n=0$ pressure deformation.
At $t=0t_{A}$, the middle-$n$ modes corresponding to the RDBMs have dominant 
contribution to \revVI{the} $n=0$ pressure deformation. 
At $t=50t_{A}$ and $t=103t_{A}$, a wide range of resonant modes with the 
down-shifted middle-$n$ mode have positive \revVI{contributions} to 
\revVI{the} $n=0$ pressure deformation. 
These results indicate that the first crash is triggered by the RDBMs and 
the second crash by the down-shifted middle-$n$ turbulence rather than 
low-$n$ modes.
The low-$n$ modes \revVI{have} stabilization effects on \revVI{the} ELM crash.

\begin{figure}
 \includegraphics[width=0.5\linewidth]{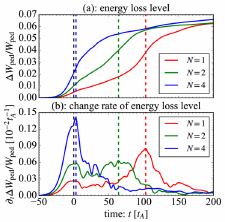}
 \caption{time evolution of (a) energy loss level and (b) its change rate 
 for the \revVI{$N=1$ case (red)}, the \revVI{$N=2$ case (green)}, 
 and the \revVI{$N=4$ case (blue), respectively}. 
 Here the time labels are set to be $t=0t_{A}$ at the first crash 
 and the timings of the second crash are highlighted with colored dashed 
 lines.}  
 \label{fig:15} 
\end{figure}

Finally the impact of toroidal wedge numbers on the two-stage crash 
process is investigated for $N=1,2,4$.
Figure~\ref{fig:15} shows the impact of toroidal wedge numbers
on the energy loss, where the result with $N=1$ is also 
plotted as a reference (see Fig.~\ref{fig:03}).
It is found that the interval between the first and second crashes becomes 
shorter with the larger toroidal wedge number, namely $\Delta t=103t_{A}$ 
in the full torus case with $N=1$, $\Delta t=64t_{A}$ in the half torus case 
with $N=2$, and $\Delta t=3t_{A}$ in the quarter torus case with $N=4$. 
In the $N=4$ case where the two crashes occur at almost same timing, the RDBMs 
directly trigger \revVI{the} ELM crash, which is consistent with our previous 
work with the $1/5$-th torus wedge torus~\cite{Seto:CPP2020}.  

\begin{figure}
 \includegraphics[width=0.95\linewidth]{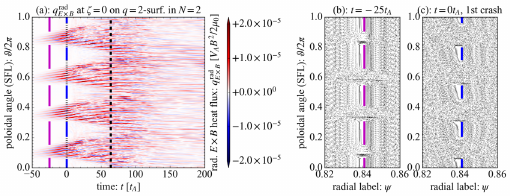}
 \caption{(a): spatio-temporal structure of radial heat flux by $E\times B$ 
 flow on the $q=2$ flux surface, Poincare plot of magnetic field line 
 including all toroidal mode component magnetic perturbation at 
 (b): $t=-25t_{A}$ and (c): $t=0t_{A}$ in the half torus case $N=2$. 
 Here the magenta lines are regions where magnetic islands exist 
 at $t=-25t_{A}$ and the blue lines are those at $t=0t_{A}$.}
 \label{fig:16} 
\end{figure}

\begin{figure}
 \includegraphics[width=0.95\linewidth]{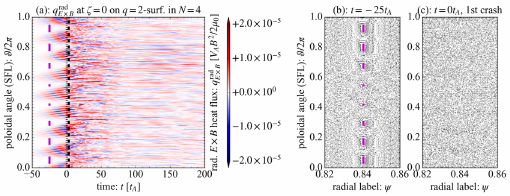}
 \caption{ (a): spatio-temporal structure of radial heat flux by $E\times B$ 
 flow on the $q=2$ flux surface, Poincare plot of magnetic field line 
 including all toroidal mode component magnetic perturbation 
 at (b): $t=-25t_{A}$ and (c) $t=0t_{A}$ in the quarter torus case $N=4$. 
 Here the magenta lines are regions where magnetic islands exist at 
 $t=-25t_{A}$.}
 \label{fig:17} 
\end{figure}

To understand the reason why the toroidal wedge number has the impact on the 
crash process, we analyze spatio-temporal \revVI{structure} of the radial 
\revVI{$E\times B$} convective heat flux $q^{\rm rad}_{E\times B}$ at the 
$\zeta=0$ line on the $q=2$ flux surface and \revVI{time evolution} of 
\revVI{the} magnetic field line topology, where the results in the $N=2$ case 
are shown in Fig.~\ref{fig:16} and those in the $N=4$ torus case are in 
Fig.~\ref{fig:17}, respectively.
In the $N=2$ case, the $m/n=4/2$ magnetic islands are generated in the early 
nonlinear phase ($t=-25t_{A}$). 
These magnetic islands have finer structure with more X-points and 
\revVI{disappear} faster compared with the $m/n=2/1$ magnetic islands 
in the $N=1$ case due to the magnetic island overlapping with higher 
harmonics, which leads to \revVI{the faster second crash}.
In the $N=4$ case, the second crash occurs more rapidly.
The $m/n=8/4$ magnetic islands are generated via nonlinear couplings in the 
early nonlinear phase \revVI{($t=-25t_{A}$)}, but they are already hidden in 
the sea of the stochasticity at the first crash.
This is the reason why the two crashes occur almost simultaneously 
in \revVI{the $N=4$ case}. 
These results indicate that the toroidal wedge number affects 
the trigger dynamics of ELM crash via nonlinear interaction.


\section{Summary and Discussions}
\label{sec:4}

In order to understand the crash mechanism of \revVI{the} RDBM-driven ELM,
we have conducted \revVI{the} analyses on simulation data in the full torus 
domain and compared with those in half and quarter torus domains.

In the early nonlinear phase, the first crash is triggered by the linearly 
unstable RDBMs and the $m/n = 2/1$ magnetic islands are nonlinearly excited 
by the parity mixing~\cite{Sato:POP2017,Ishizawa:PPCF2019} via nonlinear 
couplings of the RDBMs.
Simultaneously, \revVI{the middle-$n$} RDBM turbulence develops but is 
poloidally localized around X-points of the magnetic islands, 
leading to the small energy loss. 
The $m/n = 2/1$ magnetic islands \revVI{have} the stabilization effect 
on the RDBM-driven ELM crash.

The second crash then occurs in the late nonlinear phase. 
The \revVI{higher harmonics of $m/n=2/1$} magnetic islands are also excited 
around the $q=2$ surface via nonlinear couplings among the middle-$n$ 
turbulence. 
Since the turbulence develops from \revVI{some of their} X-points, 
it expands out poloidally. 
The second crash is triggered by the middle-$n$ turbulence 
when the turbulence covers the whole poloidal region. 

Finally, the scan of toroidal wedge number $N$ has revealed that the interval 
between \revVI{the first and second crashes} becomes shorter with the larger 
toroidal wedge number, and the two crashes occur almost simultaneously in the 
quarter torus case \revVI{with $N=4$}.
This is because that the finer magnetic islands with $m/n=2N/N$ are generated 
in the early nonlinear phase and the magnetic stochastisation occurs faster 
for \revVI{the} larger toroidal wedge number.
These results indicate that nonlinear interactions of all channels in the 
full torus domain can significantly \revVI{affect} the trigger dynamics of 
ELMs driven by the RDBMs.

It should be noted again that the linearized parallel heat diffusion term 
and the constant heat diffusivity and hyper resistivity have been employed 
in this work. 
The anomalous heat transport~\cite{Rechester:PRL1978} and 
\revVI{hyper resistivity~\cite{Kaw:PRL1979}} by magnetic stochastisation 
have not been taken into account. 
With these anomalous effects, a hyper-exponential growth of the low-$n$ 
fluctuations~\cite{Itoh:PPCF1998} and an increase of the energy loss in the 
stochastic regions~\cite{Rhee:NF2015} may be expected. 
It is left for a future work.

Within the context of the full torus case, it becomes evident that 
the nonlinear coupling between modes during the \revVI{first} crash phase 
assumes a pivotal role. 
This coupling not only leads to the deformation of the magnetic field 
configuration but also results in the disruption of magnetic 
flux surfaces through reconnections. 
Furthermore, it facilitates the conversion of \revVI{the} perpendicular 
kinetic energy into \revVI{the} magnetic energy, effectively postponing 
the substantial loss of \revVI{the} internal energy.

It's imperative to underscore that extensive internal energy loss, 
large ELM crashes, and significant inter-energy transport predominantly 
arise from the intricate overlaps of fine magnetic islands, rather than 
the mere presence of low-$n$ magnetic islands. 
Low-$n$ islands still maintain a relatively modest ratio between the volume 
of stochastic regions near the X-points and the confined regions inside the 
island. 
It is within these stochastic regions that substantial internal energy 
transport takes place.

As the number of fine islands continues to increase and they begin to 
overlap, this ratio escalates, leading to heightened internal energy 
transport and ultimately culminating in the occurrence of large ELM crashes. 
This dynamic likely accounts for the observed variation 
in crash behavior for different toroidal wedge numbers, 
while the ultimate size of the ELM remains relatively consistent.

\section*{Acknowledgment}
One of the authors (H.S.) would like to thank Drs. B. Zhu and N. Li for 
fruitful discussions and comments. 
This work is partly supported by JSPS KAKENHI Grant No. \revVI{20K14448} and 
performed under EU-Japan BA collaboration, under JIFT collaboration between 
National Institutes for Quantum Science and Technology (QST) and Lawrence 
Livermore National Laboratory (LLNL) and under the auspices 
of U.S. Department of Energy by LLNL under Contract No. DE-AC52-07NA27344. 
The computations were carried out on JFRS-1 supercomputer at QST and SGI 
HPE8600 supercomputer at QST and Japan Atomic Energy Agency. 
\nocite{*}
\bibliography{ref_seto_pop2023}
\end{document}